\input {epsf}
\documentclass[12 pt]{article}
\begin{document}
\begin{flushright} ULB--TH--99/21 \\hep-th/9911185\\
November 1999\\
\end{flushright}
\vspace{.3cm}

\begin{center} {\large PRIMORDIAL INFLATION\footnote{Presented
at the Erice International School on ``Basics and Highlights in
Fundamental physics'', August 1999.}}\\
\vspace{1cm}
{ F.~Englert}\footnote{ E-mail : fenglert@ulb.ac.be}\\
\vspace{.4cm}
{\it Service de Physique Th\'eorique}\\ {\it Universit\'e Libre de
Bruxelles, Campus Plaine, C.P.225}\\ {\it Boulevard du Triomphe,
B-1050 Bruxelles, Belgium}\\ {\it and}\\ {\it Department of
Physics and Astronomy}\\ {\it  University of South Carolina,
Columbia Campus}\\ {\it  South Main Street, Columbia SC 29208,
U.S.A.}\\
\end{center}
\vspace{.3cm}

\begin{abstract} A macroscopic universe may emerge naturally
from a Planck cell fluctuation by unfolding  through a stage of
exponential expansion towards a homogeneous cosmological
background. Such primordial inflation requires a large and
presumably infinite degeneracy at the Planck scale,  rooted in  the
unbounded
 negative gravitational energy stored in conformal classes. This
complex Planck structure is   consistent with a quantum tunneling
description of the transition from the Planck scale to the
inflationary era and implies, in the limit of vanishing Planck size,
the Hartle-Hawking no-time boundary condition. On the other hand,
string theory  give credence to the holographic principle and the
concomitant  depletion of states at the Planck scale. The apparent
incompatibility of primordial inflation with holography  either
invalidates one of these two notions or  relegates the nature of the
Planck size  outside the realm of quantum physics, as we know it.
\end{abstract}
\begin{verse}

\hspace {2.5cm}\footnotesize{\it There are more things in heaven
and earth, Horatio,\\
\hspace {2.5cm}Than are dreamt of in your philosophy.\\
 \hfill Shakespeare (Hamlet, Act I Scene V)}
\end{verse}

\addtocounter{footnote}{-2}
\newpage
\noindent {1. INTRODUCTION}

\bigskip

Primordial inflation is a mechanism whereby  {\em a}  universe
emerges from  a Planckian fluctuation  of gravity and matter and is
stabilized by  an exponential expansion of the cosmological
background\footnote {Talking about {\em a} universe may seem
contradictory.  The reason for this terminology is that one has
often the prejudice that {\em the} universe, comprising everything,
is endowed with a global arrow of time, but there is no theoretical
indication that  such an entity does exist. I shall  call {\em a}
universe (or a universe-like configuration) any space-time
configuration with a well-defined  arrow of time containing a
large number of Planck cells,  whether or not such structure
resembles {\em our} universe.}.

The primordial inflationary scenario  arose from  an attempt to
understand in scientific terms the problem posed by the birth of
our universe  and by the homogeneity of its huge cosmological
background. The idea that our universe originated from an energy
conserving  quantum fluctuation  was first proposed by Tryon \cite
{tryon}. We used  related ideas to search for a mechanism which
does not require a fine tuning of the initial conditions for the
cosmological expansion \cite {BEG, CE}.

When extrapolated backwards in time  the adiabatic expansion
approaches one Planck time of the classical singularity,  the
presently observable universe  still comprises about
$10^{87}$ Planck cells. To avoid fine tuning, such a huge spatial
extension requires the existence of a preadiabatic phase in which
the cosmological background can develop causally from an initial
Planckian cell fluctuation\footnote{I am assuming here and in
what follows the validity of classical general relativity up to the
Planck scale.}.  The basic ingredient which renders such an
evolution possible is the fact that in general relativity the total
Hamiltonian of {\em the} universe should be zero, a general feature
which is expected to hold in any future development of gravity
theory preserving invariance under time reparametrization. This
feature renders possible the creation of positive matter energy out
of the negative energy stored in  conformally flat space-times,
such as  space-times describing  homogeneous isotropic
cosmological backgrounds.  An expanding universe could then
emerge,  without cost of energy, from empty flat space-time. The
requirement that this phenomenon be localized to a Planck cell
implies that universe-like configurations  can occur within a
universe
\cite{CE,E}.

Semi-classical considerations indicate that the  cosmological
expansion needed to generate matter energy in a self-consistent
way is the exponential expansion  of a de Sitter space-time \cite
{BEG}.  Nevertheless, conventional classical general relativity
cannot yield a satisfactory   theory of a nascent universe.
Extending classical general relativity   to the Euclidean section, it
appears possible to interpret its birth  as a tunneling process. In
the limit of  vanishing Planck size ($\hbar \to 0$) one recovers in
this way the no-time Hartle-Hawking boundary condition
\cite{HH}.  More precisely the no-time boundary appears as the
limit of a thermal state where the original Planck cell has infinite
temperature\footnote{As will be shown in Section 3, this result
follows from the tunneling approach of reference
\cite {CE2}. The preheating  of  matter was anticipated in the
references \cite {BHW}.}.   Planck cells appear as very complex
structures containing a huge and perhaps an infinite number of
degrees of freedom.

Primordial inflation explains  the large scale of a universe, the
flatness and the homogeneity of the background. It also raises the
possibility of creating  the entropy of a universe during the
inflationary phase itself as internal entropy. The latter  would  be
released  as radiation entropy during the turnover to the adiabatic
era \cite {CE, E}. Such scenario would be  different from the
reheating process envisaged in more mundane inflation
mechanisms where the entropy is directly formed at turnover \cite
{review}.  Whether  this alternative would affect the  seeding  of
large scale homogeneities as usually predicted by inflation
remains to be seen,  but  is outside the scope of the present
considerations.

The huge number of degrees of freedom needed at the Planck scale
to accommodate primordial inflation is a feature corroborated   by
semi-classical considerations on black hole evaporation.  But it is
at odd with the holographic principle which states that the number
of degrees of freedom within a given space volume is limited by
the number of Planck cells of a surface bounding the volume \cite
{HS}. The holographic principle  appears to be encoded in the string
and M-theory approach to quantum gravity and implies a depletion
of quantum states at the Planck scale. The incompatibility of
primordial inflation with holography  seems to invalidate one of
these two notions. Both  could only survive  if the complex
structure of the Planck size would be outside  the  realm of
quantum physics, as we know it. Recently, 't Hooft advocated that
quantum physics  gives a statistical description of an underlying
deterministic and dissipative theory by coalescing fundamental
states into equivalence classes which define the quantum states
\cite{thooft}. The merging of all Planckian degrees of freedom into
a  single class or into few classes would indeed reconcile, in the
present context, holography with a complex structure of the Planck
size. But  no light is shed here on whether Planck size physics
should be  deterministic or not.

I present a critical review of the semi-classical approach to
primordial inflation in Section 2 and  I discuss the  tunneling
approach in section 3.  Conclusions are stated in section 4 where
the  nature of  Planck cells  resulting from these considerations
are confronted  with  M-theory arguments.

\bigskip\bigskip

\noindent{2. THE SEMI-CLASSICAL APPROACH}
\bigskip

The classical action for gravity and matter is
\begin{eqnarray}
 S&=& S_{gravity} + S_{matter}\cr &=&-{1\over 16\pi G}\int
\sqrt{-g}\,  R(g_{\mu \nu }) +
\int
\sqrt{-g}\left\{ {\cal L} (\psi_i ,g_{\mu \nu}) -
\Lambda  \right\}
\label{action}
\end{eqnarray}
 where $\psi_i$ designate matter fields. The cosmological
constant  term
$\Lambda$ is included in the matter lagrangian density
${\cal L}$.

The invariance of the action Eq.(\ref{action}) under time
reparametrization leads  to a constraint equation expressing the
vanishing of the total Hamiltonian density. If, as is reasonable to
assume for the whole universe, no boundary term   contributes to
the total energy, the total Hamiltonian $H$ satisfies
\begin{equation}
 H \equiv H_{gravity} + H_{matter} =0 \ ,
\label{constraint}
\end{equation}  expressing the vanishing of the total energy. As the
matter energy (assuming the net cosmological constant to be non
negative) is positive definite the total gravitational energy must
be negative if matter is present. To understand the origin of this
negative energy let us perform the conformal transformation
\begin{equation}
\tilde g_{\mu \nu} = g_{\mu \nu}  \exp {(-2\phi)}
\label {conformal}
\end{equation}  and rewrite the gravitational action $ S_{gravity}$
as
\begin{eqnarray} S_{gravity} = &-&{1\over 16\pi G}\int
\sqrt{-\tilde g}\exp {(2\phi)} R(\tilde g_{\mu\nu })
\nonumber\\ &-& {3\over 8\pi G}\int
\sqrt{-\tilde g}\exp {(2\phi)} \tilde g^{\mu\nu}
\partial_\mu {\phi}\partial_\nu {\phi} \ .
\label{gravityconformal}
\end{eqnarray}

Conformal classes are defined by a reference metric $\tilde g_{\mu
\nu}$ and the field  $\phi$. The ``kinetic energy''  of the
$\phi$-field is negative definite in the reference metric. The
importance of this observation for cosmology is that all
Robertson-Walker geometries describing homogeneous
cosmological expansion are conformally flat. They can be described
by a reference Minkowski space in which   the constraint
Eq.(\ref{constraint}) follows from the the energy-momentum
tensor in flat space-time deduced from
Eqs.(\ref{gravityconformal})
\begin{equation}
\tilde T_{\mu \nu}^{gravity} + \tilde T_{\mu
\nu}^{matter} = 0 \ ,
\label{conservation}
\end{equation}  with
\begin{eqnarray} \tilde T_{\mu \nu}^{gravity}= &-&{3\over 4\pi
G}[\exp {(2\phi)} \partial_\mu {\phi}\partial_\nu {\phi} -{1\over2}
\tilde g_{\mu\nu} \exp {(2\phi)}
\tilde g^{\sigma\tau} \partial_\sigma {\phi}\partial_\tau
{\phi}\nonumber\\
 &-& {1\over6} (\Delta_\mu \partial_\nu -\tilde
g_{\mu\nu}\nabla^2)\exp {(2\phi)}] \ .
\label{negative}
\end{eqnarray}

Eqs.(\ref{conservation}) and (\ref{negative}) give the usual
description of the cosmological evolution of homogeneous  matter
in a Robertson-Walker metric.  To make this explicit, let us write
the Robertson-Walker metric
\begin{equation}
 ds^2= dt^2 - a^2(t) d \Sigma^2 =  a^2(\eta) [d \eta^2 - d
\Sigma^2]
\label{geometry}
\end{equation}  where $d \Sigma^2 $ describes a reference sphere,
hyperboloid or flat Euclidean 3-space.  $\eta$ is the conformal
time   defined by $dt =  a(\eta) d\eta$.

Consider first  the spatially flat case.  Taking
$\tilde g_{\mu\nu} = \eta_{\mu\nu}=(+1,-1,-1,-1)$    and choosing
the conformal factor to be
\begin{equation}
\exp {\phi (\eta)} =  a(\eta)\ ,
\label{scale}
\end{equation}  one  can write Eqs.(\ref {conservation}) and
(\ref{negative})  as :
\begin{equation}
 -{1\over2}\left({d a(\eta)\over d \eta}\right)^2 + {4\pi G
\over 3}  a^4(\eta)\sigma =0\ ,\quad d(\sigma a^3)=-p d(a^3)\ ,
\label{flat}
\end{equation}
 where $\sigma = T^{t}_{t}=  a^{-4} \tilde T_{\eta\eta}^{matter}$ is
the matter energy density in the comoving frame and $p$ its
pressure. Eq.(\ref{flat}) is the usual Einstein equation for the
spatially flat universe written in a way which illustrates the
negative ``kinetic energy'' carried by the conformal factor.
Similarly, with different relations between $a$ and $\exp {\phi}$,
one  recovers
\begin{equation}
 -{1\over2}\left({d  a(\eta)\over d \eta}\right)^2 \mp {
a^2(\eta)\over\rho^2} + {4\pi G
\over 3}  a^4(\eta) \sigma =0
\label{curved}
\end{equation} with the minus (plus) sign for positive (negative)
spatial curvature. $\rho$ is the radius of the reference sphere or
hyperboloid.

In absence of matter the only solution to Eqs.(\ref{conservation})
and (\ref{negative}) is  flat space-time if the vacuum energy is
canceled by the
$\Lambda$ term in Eq.(\ref{action})\footnote{This cancellation is
imposed here on phenomenological ground. Note that if the
existence of a small value for the cosmological constant in our
universe is confirmed, one should  not perform a complete
cancellation and the flat space-time solution would be replaced by
a de Sitter space-time of large radius. }. Due to the negative sign
in the gravitational energy tensor Eq.(\ref {negative}), one may
envisage a energy conserving transition from empty flat
space-time to a cooperative phenomenon whereby positive matter
energy  drives an expansion which in turn creates more matter
energy.  Fluctuations around  the cosmological background will of
course provoke a departure from conformally flat metrics but the
expansion energy will remain a driving force encoded in the
negative energy stored in conformal classes.

When matter is described by a scalar field with non vanishing
expectation value, the cooperative process envisaged here is
simply the mundane inflation mechanism : matter energy is
produced at constant density as an effective  cosmological
constant resulting from some classical scalar field equation. If
the  original transition  from zero to a finite effective
cosmological constant could, in such models, originate from
gravity-matter interactions, they would  yield a primordial
inflation.

I shall now review critically a model which explores, in a
semi-classical approximation, the possibility that gravity  would
generate  the effective cosmological constant from empty flat
space \cite {BEG, E}.

Consider   a  massive scalar field interacting only with gravity.
The matter action is taken to be
\begin{equation} S_{matter} = \int
\sqrt{-g}\left\{\left[{1\over 2} g^{\mu\nu} \partial_\mu
\psi\partial_\nu\psi - {1\over 2}(M-{R\over 6})
\psi^2\right] - \Lambda
\right\}.
\label {matter}
\end{equation} The inclusion of  the $R/6$ term restores, when $M
\to 0$  classical conformal invariance ; it ensure that the number
of massive particle produced by the cosmological expansion
remains finite \cite{GMM}.

 At some initial time
$t=\eta=0$ the geometry is taken to be  empty flat Minkowskian
space-time. Assume provisionally that  after  a time $t> t_0$,
where $t_0$ is somewhat larger than
 the Planck time, space-time can be described by a non trivial
conformally flat metric $g_{\mu\nu}= a(\eta)
\eta_{\mu\nu}$. In  the reference frame
$\tilde g_{\mu\nu} =\eta_{\mu\nu}$, the normal modes  of the
rescaled scalar field $\tilde \psi =\psi \ a(\eta)$ acquire
time-dependent frequencies
$\omega_p = (p^2 + M^2  a^2(\eta))^{1/2}$.  Expanding
$\tilde \psi$ in
 creation and destruction operators
$\alpha^+(p,0),\alpha(p,0) $  at time $t=\eta=0$, one defines  the
Heisenberg state of the universe
$\vert\Omega\rangle$ by
$\alpha(p,0)\vert\Omega\rangle =0$.   As the universe expands, the
time dependent normal modes  get populated with a density $n_p(t)
=
\langle\alpha^+(p,t)\,\alpha(p,t)\rangle$. One  looks for a self
consistent solution for the gravity-matter system in the
semi-classical limit where matter is treated quantum
mechanically but gravity remains classical.

Trying, at sufficiently large time $t$, the de Sitter space-time
solution given by
\begin{equation} a(t) = \exp (t/\tau)\ ,
\label{exponential}
\end{equation}
 or equivalently
\begin{equation}
 a(\eta) =\tau / (\tau - \eta)\ ,
\label{expoconformal}
\end{equation} one writes  the proper energy density due to the
created  particles as
\begin{equation}
\sigma =  a^{-4}(t) \int_0^\infty {p^2 d p \over 2\pi^2} n_p(t)
\omega_p(t)\ .
\label{mattercreation}
\end{equation} Here, $\tau$  is the radius of the de Sitter
space-time which is taken to be of order $t_0$. The  zero point
energy in Eq.(\ref{mattercreation}) has been subtracted to ensure
the  vanishing of the  cosmological constant in flat space-time.

One can show \cite {BEG, E} by computing
$n_p(t)$ that
$\sigma$ given by Eq.(\ref{mattercreation}) becomes independent
of $t$ after a time of order $\tau$ and thus that the created
matter energy density is, at sufficiently large time $t$, constant
in time. In addition, one can verify that
$p= -\sigma$, in accordance with energy conservation for creation
of a {\em constant} energy density. The integral
Eq.(\ref{mattercreation}) can be performed analytically to yield
\begin{equation}
\sigma = {M^4 \over 64 \pi^2} [ \Psi({3\over2} + \nu) +
\Psi({3\over2}  - \nu) - 2 \ln M \tau ]
\label {density}
\end{equation} where $\Psi$ is the digamma function and
$\nu^2 =1/4 - (M\tau)^2 $ \cite{S}.  From the asymptotic form of
the digamma functions one gets  for $M\tau \gg 1$
\begin{equation}
\sigma \to { M^2\over 96 \pi^2 \tau^2}\ .
\label {limit}
\end{equation} There is an additive correction to Eq.(\ref{density})
revealed by neighboring non conformally flat metrics,  namely  the
trace anomaly
\begin{equation}
\sigma_{anomaly} = {1\over 960 \pi^2 \tau^4}\ .
\label{anomaly}
\end{equation} Clearly the anomaly contribution does not affect
the asymptotic value of
$\sigma$ given in Eq.(\ref{limit}).

\bigskip \bigskip

 \quad\epsfbox {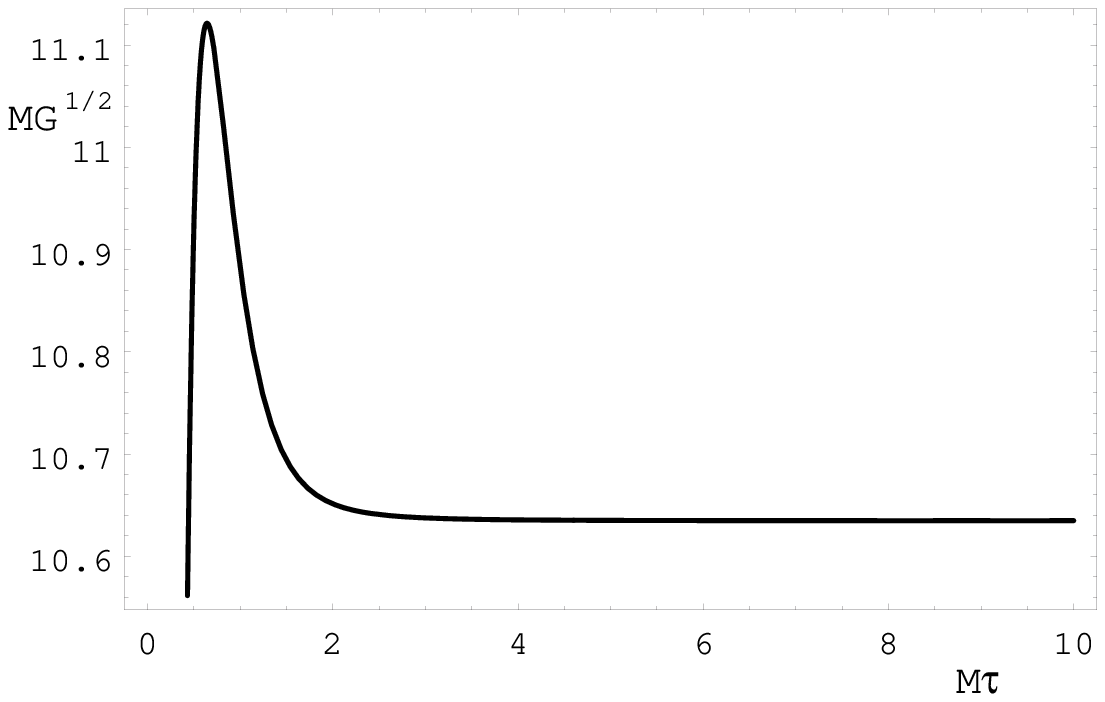}

\begin{center} {\footnotesize Fig.1 $M\sqrt{G}$ as a function of
$M\tau$. The asymptotic value of
$M$ is $6 \sqrt {\pi/G}$.}
\end{center}

\bigskip

Equating the total contribution to
$\sigma$ due to quantum effects, as given by Eq.(\ref{density}) and
Eq.(\ref{anomaly}),   to its value for the de Sitter  solution of
Einstein equation
\begin{equation}
\sigma_{de  Sitter} = {3\over 8\pi G \tau^2}\ ,
\label{desitter}
\end{equation} one obtains a relation between $\tau$ and $M$ :
\begin{equation} G^{1/2}M = \left\{ {(M\tau)^2\over
24\pi}[\Psi({3\over2} + \nu) +
\Psi({3\over2}  - \nu) - 2 \ln M \tau  +{1\over 15 (M\tau)^4} ]
\right\}^{-{1\over2}} .
\label{solution}
\end{equation}  This equation was first solved numerically in
reference \cite{S2}.  The solution  is depicted in Fig.1.  One sees
that for
$M\tau>1$ a self-consistent solution of the equations of motion
exists only if $M$ exceeds the critical value obtained by equating
the asymptotic value Eq.(\ref{limit}) to the de Sitter solution
Eq.(\ref{desitter}). The critical mass is
\begin{equation} M_{critical} = 6 \sqrt{\pi\over G}\ .
\label{critical}
\end{equation}
$M$ remains very close to its critical  value in the region
$M\tau >1$. In this region  the Compton wavelength of the particle
is small compared to the de Sitter radius and, as
$\tau$ should be greater than $\sqrt{G}$,  I shall consider only this
region. The drop towards zero below  $M\tau = 1$ is due to the
trace anomaly  which has a negligible effect for
$M\tau >1$.

It thus appears at first sight that an exponentially expanding
universe, originating from some fluctuation in empty flat
space-time,  can be sustained through creation of particles of
mass  close to the critical mass. But there is a catch. The
asymptotic value of
$\sigma$ in Eq.(\ref{limit})  yields a trace of the
energy-momentum tensor proportional to the curvature and one
could absorb it in a renormalization of the gravitational coupling
constant. This would yield  a renormalized matter density
$\sigma_{ren}$
\begin{equation}
\sigma_{ren} = {M^4 \over 64 \pi^2} [ \Psi({3\over2} + \nu) +
\Psi({3\over2}  -
\nu) - 2 \ln M \tau  -  { 2 \over3 (M \tau)^2} +{1\over 15 (M\tau)^4} ]
\label{renormalized}
\end{equation} and a renormalized coupling constant $ G_{ren}$
\begin{equation}
 G_{ren}= {G \over 1- GM^2/ 36 \pi}\ .
\label {ren}
\end{equation} These are   well known results \cite{spindel}.
Clearly, the renormalization of the gravitational constant ensures
that the subtraction in the renormalized energy density does not
affect the solution,  expressed in terms of the bare $G$,
represented in Fig.1.  But what goes wrong is the renormalization
procedure for $M$ greater than
$M_{critical}$ : the renormalized gravitational coupling becomes
negative  precisely at the lowest mass value for which  a  solution
to the semi-classical equations  exists.

The breakdown of the renormalization procedure indicates that
new effects must be taken into account in the
gravitational-matter system. This is not astonishing. The  high
value of the critical mass suggests  that strong coupling effects
appear in the  system around this value. These could lead to   the
formation of black holes \cite {CE}, or, if new matter degrees of
freedom are introduced, to extended objects such as ``strings''
\cite {AEO} .  The relation between unrenormalized quantities
depicted in Fig 1  might still  describe approximately the
formation of the massive objects but the scalar field should be
viewed only as a phenomenological description of such  objects
\cite{CE}.  An interesting possibility is that, in contradistinction
with more mundane inflation scenario, these objects would carry
internal entropy, (as   black holes or strings do), which could bring
them in (unstable?) thermodynamic equilibrium with the de Sitter
background. The conversion of internal entropy, created
exponentially during inflation, into radiation entropy at the
turnover to the adiabatic era would then be the source of the
cosmic background radiation\footnote{The subtracted term  in the
renormalized energy density Eq.(\ref{renormalized}) includes the
dominant contribution to the number of quanta created for $
M\tau>1$. Such  subtraction can be done because the number of
quanta of  local field excitations is not an invariant in general
relativity, but it becomes questionable in a phenomenological
description.}.

Up to now, I have considered a spatially flat section of a de Sitter
cosmological background  emerging from flat space-time.  To avoid
fine tuning, the original  fluctuation should be confined to a Planck
size. As the inflation renders in a few Planck times the initial
spatial curvature  negligible, one is justified to  consider simply
spatially flat cosmological backgrounds. However  the emergence
of the universe should appear as a  process localized on its Planck
scale and not as a global process in an underlying flat background.
To visualize how this could happen, choose  a coordinate system in
the flat background such that the time coordinate coincides with
the conformal time
$\eta$ of the emerging universe. Namely
 define a local conformal time
$(\eta,{\bf x})$ such that
\begin{equation}
 ds^2= a^2(\eta,{\bf x}) (d\eta^2 - d{\bf x}^2)
\label{baby}
\end{equation}
 where the ${\bf x}$ dependence of the scale factor is introduced
to parametrize the transition region : $ a=1$ well outside the
Planckian region $\vert {\bf x}\vert \gg  \sqrt G$ and $ a=
a(\eta)$ for $\vert {\bf x}\vert \leq t_0$ where it describes at
some initial time
$\eta = 0$ the onset of a  de Sitter universe of radius
$\tau\simeq t_0$. Explicitly one has Eq.(\ref{expoconformal}) for
$\eta > 0$ and thus an exponential expansion in the proper time $t$.

Thus, in the ``external'' time $\eta$ the creation process and the
whole de Sitter period takes only  a time of order $\tau$. Although
the universe has finite spatial extension of order
$\tau$ in $\vert {\bf x} \vert $, an internal observer will
contemplate an homogeneous cosmological background as long as
his visible horizon is, in these coordinates, less than
$\tau$. The appearance of the inhomogeneous   drop in scale at the
edge of the universe  would signal to the internal observer the end
of his universe. It follows from the conformal flatness of the
cosmological background that the whole history  subsequent to the
de Sitter period  is   delimited, in the conformal time
$\eta$, by the light rays emitted from the edge and therefore is
also of order $\tau$.  Hence the ``external'' observer would describe
the rise and fall of our universe as a fluctuation on a scale $\tau$
comparable to the Planck scale. The reason ``we'' can exist during
this seemingly short time span is the enormous dilation of the
proper time $t$ generated by the primordial inflation.

The possibility of generating a universe from a Planckian
fluctuation in empty flat space raises the question of the
significance of this reference space. Is it only a mathematical
device or has it physical significance? If the transition from a
Planckian event to the cooperative phenomenon does indeed occur,
the answer must be that the reference space is physical because
nothing could prevent a similar phenomenon to occur {\em within }
our universe, as, on scales large compared to the Planck scale but
small compared to cosmological scales, any background can be
viewed as flat. This does not lead to a contradiction because
universe-like configurations within our world would be  only
Planckian (or transplanckian -see below) fluctuations of  our
space-time. Universes within universes can therefore be generated
in a large (infinite?) number of different ways and universe-like
fluctuations may well be  dominant  Planckian effects. Note that
there is no reason for two such baby-universes to have a common
comoving time. It is  with respect  to their own comoving time
that their original Planck size has a fixed scale and spatial
extension. The comoving time of a baby-universe fixes locally in
the mother universe the $\eta$ variable in the metric Eq.(\ref
{baby}). In particular two baby-universes may have opposite arrows
of time.

The transition from a Planckian event to a universe-like
configurations in an underlying space-time poses a new problem
because, as mentioned above, the scale factor develops a huge
gradient at its edge.   We see from Eq.(\ref {negative}) that a
spatial gradient of the scale factor
 gives rise to a negative energy density in $\tilde
T_{\mu\nu}^{gravity}$ as does its   time derivative hitherto
considered.  Local compensation by positive  energy has to take
place, possibly by forming  new universe-like configurations. Note
that these would seem to occur at  much smaller scales than the
Planck scale but such transplanckian effects are fictitious : the
Planck scale is defined by the baby universe and is the same as
before ; it is only its parametrization in the mother universe which
makes it look different. However the generation of such
throat-universe would only further steepen the gradient of the
scale factor, thus giving birth to more and more new universes
containing themselves potentially universe-like configurations.
The space-time structure within a universe can thus be drastically
altered at the Planck scale : universe-like configurations may form
a ``foam of universes''~\cite {CE, E}. Although the foam may appear
as virtual,  the underlying degrees of freedom should exist.

In view of the complexity of the structure due to the negative
unbounded energy of the scale factor and to its necessary
compensation by positive energy,
 the transition era can probably not be described at the classical or
at the semi-classical level. If a   cosmological background
emerges  out of a Planck size,
$\hbar$ should enter the game in a fundamental way. An
approximate quantum mechanical description of the  creation
process,  taking into account   the  complex structure of its
Planckian origin,  should then be possible. Let us explore this
approach.

\bigskip
\bigskip

\noindent 3. THE TUNNELING INTERPRETATION
\bigskip

I shall assume that an approximate quantum description of the
transition period can be obtained from a tunneling of the scale
factor through a barrier from the original Planck scale to the
macroscopic inflationary period.  Although I cannot  justify   this
procedure, there is an important consistency check : the tunneling
of  the scale factor must be consistent with a high degeneracy of
states at the Planck side of the barrier and with an a priori
complete lack of information about these states. Remarkably this
will indeed be the case.

In order to describe a tunneling process, one must  specify
boundary conditions. These  will be  fixed from the requirement
that the emerging de Sitter space-time has a well defined
Hawking-Gibbons background temperature
\begin{equation}
\beta^{-1}_{GH} = 1/(2\pi \tau)
\label{hawkgibbs}
\end{equation} and an entropy $\cal S$ given, up to an integration
constant, by the quarter of the area of the event horizon \cite {HG}.
In our notations
\begin{equation} {\cal S} = {\pi \tau^2 \over G}\ .
\label{entropy}
\end{equation}

To understand how entropy and temperature determine the
boundary condition of the tunneling process, I first review how
entropy can be generated within a closed system  in an energy
eigenstate by a ``tunneling of time''
\cite {CE2}.  In such system, time evolution can only be defined by
correlations between subsystems. Let us assume the existence of
a  subsystem for which the WKB limit is valid in certain regions of
space ; this subsystem will be called the ``clock''. We shall see that
correlations of subsystems to the clock define in these regions
the  time variable \cite {BBE,CE2} but that the tunneling of the
clock can produce entropy. This mechanism will then be
generalized to our problem where the total energy is zero from the
constraint  Eq.(\ref{constraint}) and where the role of the clock
will be  played by the scale factor.

Consider a closed system where a  non relativistic  object with
one very massive degree of freedom, $x$, representing the ``clock",
is correlated to the remaining ``matter" part of the system  by
energy conservation only. No explicit interaction between clock and
matter is imposed but the argument below is easily extended to
the case of matter following adiabatically the clock.
 Labeling (unnormalized) matter eigenstate
 of energy $\epsilon_m$  by $\vert \chi_m \rangle$, each
eigenstate $\vert
\chi_m\rangle$ is correlated by quantum superposition to a clock
state vector  of energy $E-\epsilon_m$.  One writes
\begin{eqnarray} H \vert \Psi\rangle &=& (H_{clock} +
H_{matter})\vert
\Psi \rangle   = E  \vert\Psi\rangle , \nonumber\\
 H_{clock}& =& -{1\over 2M} {\partial^2\over \partial x^2} + U(x) \
,\nonumber\\
 H_{matter}\vert \chi_{m} \rangle &=& \epsilon_m
\vert \chi_m \rangle\ .
\end{eqnarray} Expanding $\vert\Psi\rangle$ in matter eigenstates
\begin{equation}
\vert\Psi\rangle =\sum_m \Phi_m(x) \vert \chi_m
\rangle\ ,
\end{equation}
 one gets
\begin{equation}
\left\{ { d^2\over dx^2} + 2M \left[ E -\epsilon_m - U(x)
\right]
\right\}\Phi_m (x) =0\ .
\label{clock}
\end{equation}
 Here $U(x)$ is some potential which vanishes as  $x
\rightarrow\pm \infty
$. I shall take all the $\epsilon_m$ positive and assume
\begin{equation}
\epsilon_m \ll\vert E-U(x)\vert \ .
\label{small}
\end{equation}

In the classically permitted regions,  $E-U(x) > 0$, the WKB
forward wave solution of Eq.(\ref{clock}) is
\begin{equation}
\Phi_m (x) = {1\over \sqrt{p_m(E_m,x)}}
\exp [i W (E_m)] \ .
\end{equation}
 $W$ is the classical Legendre transform of the classical clock
action
\begin{equation} W(E_m,x) = S + E_m T = \int^x_{x_i} p_m(E_m,x')
dx'
\label{legendre}
\end{equation}
 where  $T$ is the time  recorded by the clock since a chosen
initial position $x_i$. The clock energy $E_m$ is equal to
$E-\epsilon_m$ and  its momentum $ p_m(x,E_m)$ is
$\sqrt {2M [E_m - U(x)]}$.

Expanding $E_m$ to first order around $\epsilon_m =0$ one gets
\begin{eqnarray}
\vert\Psi\rangle &=& \Sigma_m {1\over \sqrt{ p_m(x,E_m)}} \exp
[iW(E_m, x)] \vert
\chi_m\rangle\nonumber\\ &\simeq& {1\over \sqrt{
p(x,E)}}\exp(iW(E,x))\Sigma_m \exp[ -i{\partial W(E,x)\over
\partial E}
\epsilon_m]
\vert \chi_m\rangle \nonumber \\ &\simeq& {1\over
\sqrt{ p(x,E)}}\exp (iW(E,x))\exp(-iT H_{matter})\Sigma_m
\vert
\chi_m\rangle
\label{time}
\end{eqnarray}
 where  from the classical action principle
\begin{equation}
 T(E,x)={\partial W(E,x)\over \partial E}\ .
\end{equation}

It follows from Eq.(\ref{time}) that, under the prescribed
conditions,  any matter wave function $ \vert\chi\rangle =
\Sigma_m\vert
\chi_m\rangle$ evolves according to the time dependent
Schrodinger equation with the time recorded by the  clock.

This result is valid in the WKB limit as long as backward waves
generated from the second order equation Eq.(\ref {clock}) do not
affect
$\vert\Psi\rangle$ significantly. An extreme departure of this
condition occurs if the clock has to tunnel through a barrier
between two turning points $x=a$ and $x=b$,  as shown in Fig.2 .

Let us call the regions $x<a$ and $x>b$ respectively the regions
``before'' and ``after'' tunneling.

Taking ``before''  tunneling forward waves only,  backward wave
will be generated ``after'' tunneling. In the limit of large clock
mass, coherence ``after'' tunneling between forward and backward
wave gets lost. The ratio $ N_m$ of the squared forward
amplitudes  ``after'' and ``before'' the tunneling of the clock with
energy $E_m = E -\epsilon_m$ is the corresponding probability
ratio for finding matter with energy $\epsilon_m$.
$N_m$ is  the inverse transmission coefficient through the barrier.
 For large barrier it is
\begin{equation} N_m =  \exp [W_e(E_m)]
\end{equation} where
\begin{equation} W_e(E_m)=\oint p_e(E_m,x) dx =
 \oint \sqrt{ 2M[U(x) -(E-\epsilon_m)]} dx
\end{equation} is the ``Euclidean'' continuation of the action
Eq.(\ref {legendre}) defined by
\begin{equation} W_e (E_m, U(x)) =W (-E_m, - U(x) )\ .
\label{euclidean}
\end{equation} Taking into account the condition Eq.(\ref {small}),
one gets
\begin{equation} N_m(E_m) = N_0 \exp[{\partial W_e(E)\over
\partial E}
\epsilon_m]= N_0 e^{\theta\epsilon_m}\ ,
\label{probability}
\end{equation} where $\theta$ is the Euclidean time spanned by the
clock in a round trip under the barrier, and
\begin{equation} N_0=\exp W_e(E)\ .
\end{equation}

\bigskip \bigskip

\epsfbox {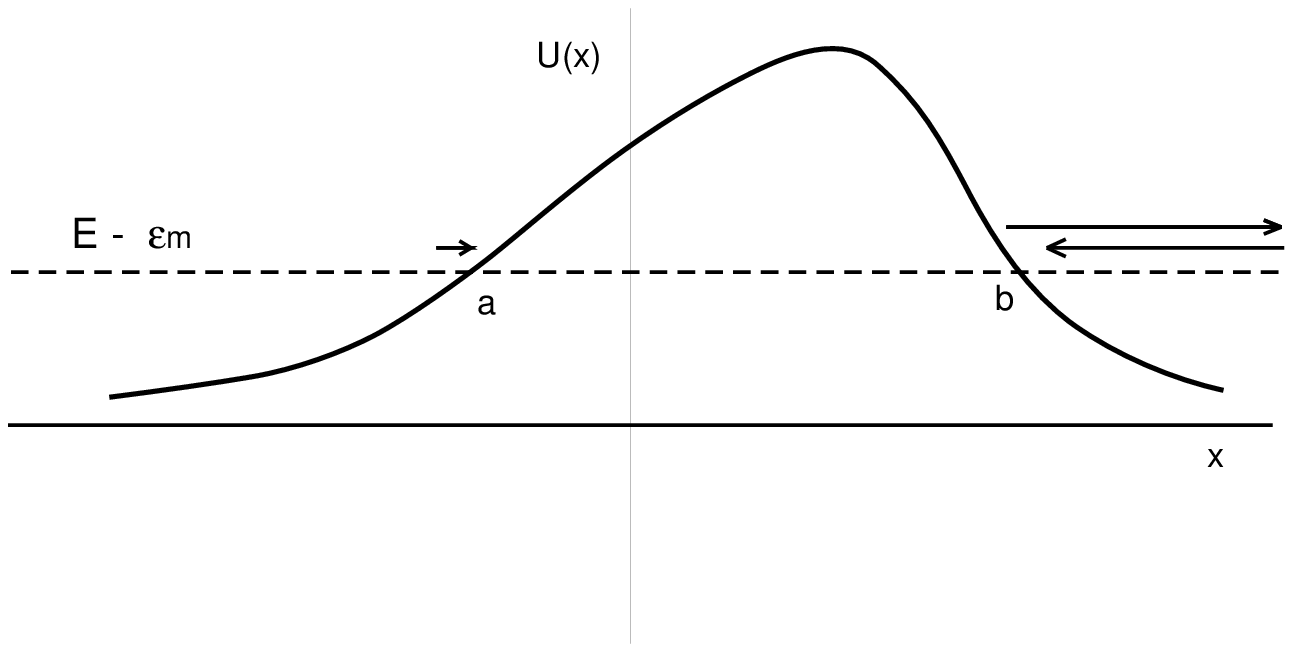}
\begin{center} {\footnotesize Fig.2 Tunneling of a clock of energy
$E-\epsilon_m$.}
\end{center}
\bigskip

The shift of sign in  Eq.(\ref {euclidean}) makes the exponent in
Eq.(\ref{probability}) positive. Thus  the probability ratios
$ N_m$ have a Boltsmann distribution with a {\em negative}
temperature
$\beta^{-1} = -\theta^{-1}$.  The temperature is negative because
higher energies for matter states mean lower energies for the the
clock and hence higher inverse transmission coefficient
 $ N_m$  for the clock. Thus higher  matter energies are favored by
the tunneling.

To assess the thermodynamic significance of  Eq.(\ref{probability})
we rewrite it more generally as
\begin{equation}
\delta W_e(E_m) = -\theta \delta E_m =\beta\delta E_m \ .
\label {wentropy}
\end{equation} This equation is again the classical action principle
but formulated for ``Euclidean'' closed trajectories. Consider
matter in equilibrium with the clock  ``before'' and ``after''
tunneling,  respectively at temperatures
$\beta^{-1}_a$ and $\beta^{-1}_b$ such that
\begin{equation}
\beta =
\beta_b - \beta_a \ .
\label {temp}
\end{equation} If ${\cal S}_b(E)$  and ${\cal S}_a(E)$ are the
corresponding clock entropies, we see that Eq.(\ref{wentropy})
implies
\begin{equation} W_e(E) = {\cal S}_b(E) -{\cal S}_a(E)\ .
\label{difference}
\end{equation}
$W_e(E)$ can be interpreted  as the entropy gained by the clock
from the backward wave generated by the tunneling process.  In the
particular case where matter is in equilibrium with the clock
``after'' tunneling at the temperature $\beta ^{-1}$, the equilibrium
temperature ``before'' tunneling must be infinite and the entropy of
the clock becomes simply  $W_e(E)$, up to an integration constant
${\cal S}_a$  independent of the clock energy.

Thus, this simple model indicates that a macroscopic clock may
gain entropy by tunneling. The above analysis may seem rather
academic because on the one hand  the model does not contain
dynamic elements to realize thermal equilibrium and  infinite
temperature and on the other hand   the    temperature $\beta$ is
negative. This is not the case. We see  from Eq.(\ref {probability})
that the sign of $\beta$ is related to the sign of the clock
Hamiltonian. Changing indeed its sign and keeping the matter
Hamiltonian positive definite would lead to
$E_m =-E +\epsilon_m$ in the  equations for
$W$ and $W_e$. Hence in Eqs.(\ref{probability}) and
(\ref{wentropy})
$\theta$ should be replaced by
$-\theta$.\footnote{ One may verify that the shift of sign does not
affect the definition of time in the classically allowed region
because for forward waves, the shift of sign in
$E_m$ is compensated by a shift of sign in the momentum which is
then opposite to the velocity.}   Negative clock energies would
 lead to positive equilibrium temperature for matter. One might
thus expect that in gravity where the ``gravitational clock'' has
negative energy and where dynamics is contained in the many
degrees of freedom, the tunneling of time should produce  entropy
and positive  temperature. I now show that this is indeed the case.

In the cosmological context, I shall, as in previous sections, retain
from gravity only the scale factor  but I now allow it to be
described quantum mechanically  in classically forbidden regions.
We shall then  see that the gravitational clock  can  tunnel from a
Planck size region to a de Sitter background of radius $\tau$. As
mentioned above, the boundary condition for the tunneling process
should be consistent with the known entropy and thermal
properties of the de Sitter background. This will determine on
which side of the tunneling  the wave function of the gravitational
clock is small and  its thermal properties on both sides.

The gravity Hamiltonian can be deduced, up to a multiplicative
factor from Eq.(\ref {flat}) or more generally from Eq.(\ref
{curved}). The mean  matter density $\sigma$, that is the
cosmological constant of the de Sitter space-time, is now included
in the gravitational Hamiltonian  As the open and flat spatial
sections of the de Sitter space do  not span the geodesically
complete manifold, I shall describe it in terms of closed spatial
sections.
  Eq.(\ref {curved}) yields
\begin{equation}
 -{1\over2} a'^2 -{1\over2} {a^2\over \rho^2} + {4\pi G\over
3}\Lambda a^4 =0
\end{equation}  where the prime denotes differentiation with
respect to the conformal time
$\eta$ and $\sigma$ has been equated to the (dynamic)
cosmological constant
$\Lambda$ of the de Sitter space-time.  To obtain the Hamiltonian
in the dimensionless conformal time $ \tilde\eta =
\eta/\rho$, the left hand side must  multiplied  by
$(3\pi /2 G) \rho^4$ to normalize correctly the (average) matter
contribution in the comoving volume $2\pi^2 \rho^3 a^3$. Defining
$ \tilde a^2 = (3\pi/ 2 G) a^2\rho^2 $ one obtains
\begin{equation}  H_{grav} = - {1\over2}p_{\tilde\eta}^2 -
{1\over2}( \tilde a^2 -
\lambda  \tilde a^4) \ ;\quad \lambda = {2G\over 3
\pi\tau^2}\ ,
\label {gravity}
\end{equation} where $p_{\tilde\eta}$ is the momentum conjugate
to
$ \tilde a$.

In absence of quantum matter, $H_{grav}=0$.
$p_{\tilde\eta}^2$ is positive for $ \tilde a >
\lambda^{-1/2}$ and reaches zero at
$ \tilde a=0$, and the latter point opens up into a finite
neighborhood in presence of positive definite energy matter. As in
the non relativistic model  one has a potential barrier, shown in
Fig.3,  separating two turning points.  The turning point at $ \tilde
a=0$  can be identified with a Planck size wormhole  and  classical
inflation sets up at the turning point $\tilde a =  \lambda^{-1/2}$.
Geometrically, the tunneling region is a half Euclidean 4-sphere of
radius $\tau$. Taking the radius of the reference
 3-sphere in Eq.(\ref{geometry}) to be $\tau$, the turning point
$\tilde a =  \lambda^{-1/2}$ is the 3-sphere $a=1$ (see Fig.4).

To get the entropy $W_e$  and the  temperature difference
$ \beta^{-1}$ of the scale factor ``clock''  it is not necessary to
include explicitly the quantum matter contribution. It suffices to
evaluate the Euclidean action
$W_e(E=0)$ between the turning points and the Euclidean time
spent under the barrier. Thus
\begin{equation}  W_e(0)= \oint p_{\tilde\eta}( \tilde a) d\tilde a
=2
\int_0^{\lambda^{-1/2}} z ( 1-
\lambda  z^2)^{1/2} d  z= {2\over3\lambda}= {\pi \tau^2
\over G}\ .
\label {dSentropy}
\end{equation}
 This coincides with the horizon entropy of de Sitter space-time
Eq.(\ref{entropy}). Thus
\begin{equation} {\cal {S}}_{de  Sitter} = W_e(0) + C
\label{entropytotal}
\end{equation} where C is an energy independent constant.

\bigskip\bigskip

\quad \epsfbox {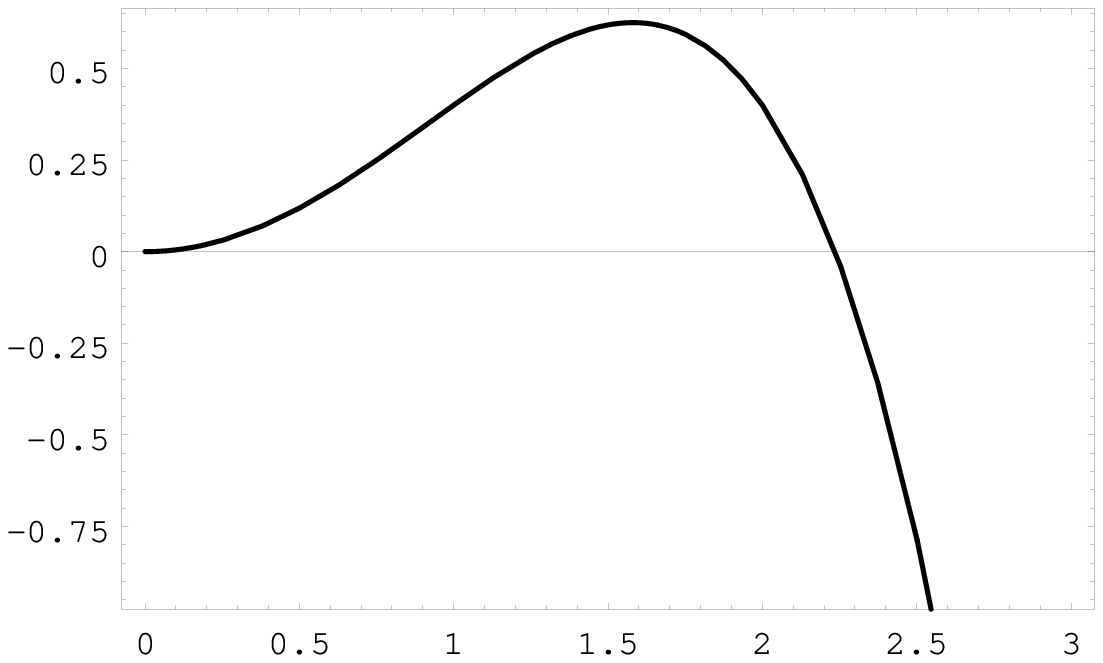}

\begin{center} {\footnotesize Fig.3  Potential $V(\tilde a) = (\tilde
a^2 - \lambda\tilde a^4)/2$ as a function  of $\tilde a$ for
$\lambda = 0.2$}
\end{center}

\bigskip

Consistency between Eq.(\ref{entropytotal}) and
Eq.(\ref{difference}) requires that the equilibrium temperature
``before'' tunneling be infinite. This can be checked more directly
by evaluating the time $\theta = \beta
$ spent under the barrier and showing that this time is equal to the
inverse equilibrium temperature of the de Sitter space-time.

The latter is the (positive) Hawking-Gibbons background
temperature Eq.(\ref{hawkgibbs}) which is the global equilibrium
temperature of de Sitter space-time with static test matter.  This
global temperature is given by the periodicity in the time
$t_s$ of the  Euclidean continuation of the
 metric of a static de Sitter patch
\begin{equation} ds^2= (1-{r^2\over \tau^2}) dt_s^2 +  (1-{r^2\over
\tau^2})^{-1} dr^2 + r^2 d\Omega^2\  .
\label{static}
\end{equation} One easily verifies that the $t_s$-period  is $2\pi
\tau$.

Eq.(\ref{static}) is the metric of the 4-sphere which can also be
described by the Euclidean continuation of the  coordinate system
for comoving observers
\begin{eqnarray} ds^2&= &dt_c^2 + a^2(t_c)  d\Sigma^2 \ ,
\nonumber\\ a(t_c)& =& \cos (t_c/\tau)\ ,
\label{comoving}
\end{eqnarray} where $d\Sigma$ is the line element of a 3-sphere
of radius
$\tau$.  The tunneling region  between $\tilde a =0$ and  $\tilde a
=\lambda^{-1/2}$, shown in Fig.3, is depicted in Fig.4 as  half  the
4-sphere of radius $\tau$ laying between the wormhole at $a=0$
and the equatorial 3-sphere
$ABCD$ at $a=1$ .

This 3-sphere, which in the comoving system Eq.(\ref{comoving})
is a slice of the 4-sphere at
$t_c=0$, is the union of two static patches : $BCD$ at time
$t_s =0$ and  $BAD$ at time $t_s =\pi$.  In the static coordinate
system, the tunneling region is spanned by the motion of the patch
$BED$  from $BCD$ to $BAD$.  The period of the round trip is
$\beta = 2\pi \tau$.  Thus
$\beta^{-1}=\beta_{GH}^{-1} $ which is the equilibrium
temperature. This confirms that the equilibrium temperature
``before'' tunneling is infinite.

\bigskip \bigskip

\  \epsfbox {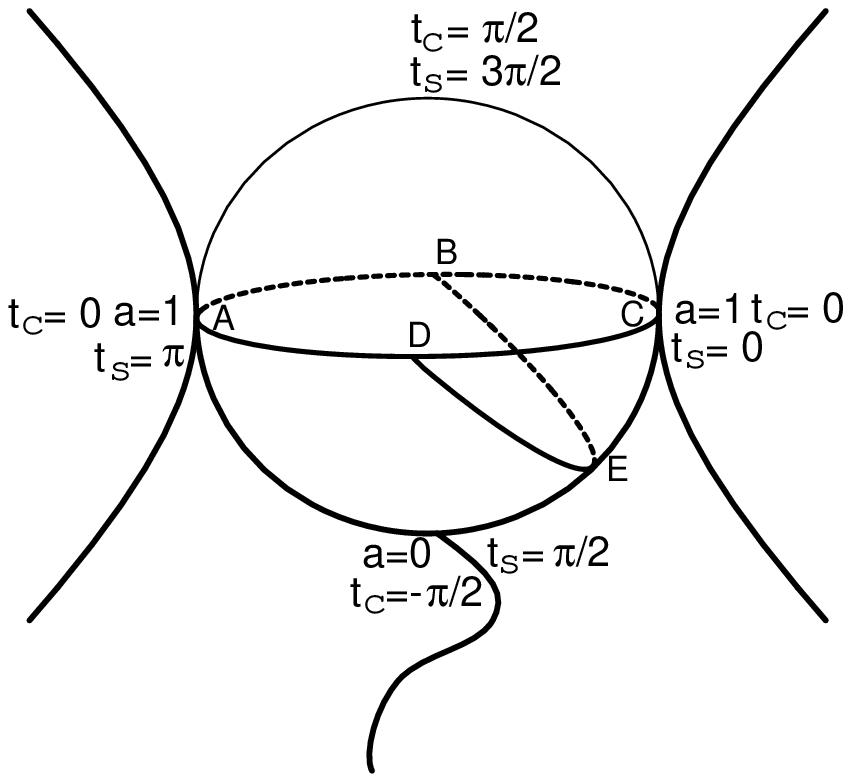}
\begin{center}
 {\footnotesize Fig.4 Tunneling through the half Euclidean sphere,
parametrized by  $\tilde a$ in Fig.3.}

\end{center}

\bigskip

The preheating of matter at infinite temperature ``before''
tunneling is consistent with the qualitative semi-classical picture
discussed in Section 2.  Namely,  an  infinite temperature, if
realized dynamically, requires a large density of states at the
Planckian origin and translates, in the scale factor
approximation,   complete ignorance about these states.

These results hinge upon the boundary conditions which must be
chosen so that {\em the wave function for the scale factor is small
``before'' tunneling and exponentially large at the onset of the
classical inflationary behavior}. In the limit of vanishing Planck
size ($\sqrt \hbar  \to 0$) such boundary conditions are equivalent
to fixing the wave function to be zero at $a=0$ in the Euclidean
section. One thus recovers the no-time  Hartle-Hawking boundary
condition \cite {HH}\footnote{In Linde's  chaotic inflation \cite
{linde}, which may  be viewed as a realization of the foam, the
boundary conditions in a tunneling interpretation are the opposite
one
\cite  {linde2}.  Namely the  wave function is exponentially small
``after'' tunneling.  This appears motivated by considerations of
creation probabilities but does not   yield the
 thermal properties of the de Sitter space-time. I want to stress
that here there is no global arrow of time for {\em the} universe
and that the tunneling operates in  the comoving time of the
nascent universe. Such tunneling cannot be interpreted as  a
probability of creation.}.

Finally we remark  that when $\tau \to 0$ the entropy gained from
tunneling disappears while the temperature of the mini-de Sitter
goes to infinity, matching the wormhole  temperature. The density
of states there must be energy independent on the scale of matter
excitations. If they could be counted as quantum mechanical
states, they would provide an integration constant to the de Sitter
entropy which may even be infinite. I now discuss this issue.

\bigskip
\bigskip
\noindent 4. HOLOGRAPHY AND THE NATURE OF  PLANCK CELLS
\bigskip

Semi-classical arguments on primordial inflation  lead to  a huge
complexity of the Planck scale and this picture is consistent with
the tunneling interpretation. The semi-classical approach to black
hole leads  to the same conclusion. Indeed, if one assumes,
according to the original derivation  \cite {hawking}, that the
Hawking radiation carries no information about the infalling
matter, one  may increase indefinitely the information loss by
sending more and more objects into the hole and letting it
evaporate \cite {ACN}.  When finally the black hole evaporates  up
to the Planck size, it would leave an infinite or at least a very
large\footnote {There might be some cut off on the matter that
could be send into the hole due for instance to a finite size
universe.} entropy. The black hole entropy would then contain, in
addition  to the horizon entropy $A/4$,  a   very large or perhaps
infinite integration constant. The latter,  like the integration
constant needed in the de Sitter entropy, would count the
degeneracy  of Planckian states.

This is not what comes out of string theory. The most remarkable
achievement of the superstrings and of M-theory considerations is
the counting of quantum states for near extremal black holes. Not
only is the value $A/4$ recovered but the integration constant is
zero \cite {all}. Also the Hawking radiation appears to contain the
necessary information to ensure unitarity for the scattering
matrix. Although no rigorous proof has been established for
Schwarzschild black holes, the very fact that it seems possible to
connect them adiabatically to near extremal ones, through
performing reversible work in upper dimensions
\cite {ER},   indicates that these conclusions should not be altered
in a fundamental way for Schwarzschild black holes. Namely  the
integration constant should remain zero or small. This  supports
the holographic principle which limits the number of quantum
degrees of freedom in a volume $L^3$ by the number of Planckian
cells on a surface of area
$L^2$ \cite {HS}. Such drastic reduction in the number of states  is
expected in general relativity if, in accordance with the second
principle of thermodynamics, the  entropy within the volume is
bounded by the area entropy of the largest black hole fitting the
volume, {\em provided the integration constant in the  entropy is
not infinite or exceedingly large}.

In the string-M-theory approach to quantum gravity,  T-duality
and   high temperature analysis \cite{rabin} also suggest a
depletion of states at the Planck scale.  Even  more direct evidence
for holography \cite {AdS}, although in a peculiar local way and in
cases which are not directly physically relevant,  follows from the
proposed correspondence between  Anti de Sitter space-time and
some Conformal Field Theory  living on its boundary,
correspondence which seems to be borne out  for some limiting
values of the parameters \cite {malda} .

We are  confronted  with a dilemma which raises fundamental
questions.  Possible alternatives are :

a) The holographic principle is true and the Planck size is
essentially empty. Primordial inflation is incorrect.  Also
incorrect are all semi-classical considerations for black hole
physics and in particular the original derivation of the Hawking
radiation. At present, the    objection to such conclusion  is mainly
of philosophical nature.  The notion of space-time, as we use it,  is
perhaps not  operative beyond the Planck scale.  But why should
physics  disappear at a
 scale which might be the cradle of our universe.  Such conclusion
would carry a strong anthropomorphic connotation which hitherto
did never improved our  ken.

b) The holographic principle is wrong and the Planck scale has a
huge or infinite degeneracy of quantum states. The string-M-theory
considerations would be totally irrelevant for physics.  But why
should a tentative approach to quantum gravity,  which led   to a
remarkable understanding of the entropy of at least some black
holes in terms of a counting of quantum states, be completely void
of physical significance.

c) The holographic principle is true and so is  the complexity  of a
Planck cell,  but usual quantum mechanics is not operative beyond
the Planck scale. This could mean that the Planck energy scale
marks the scale {\em below} which the quantum description of the
world enters the physical description. The appearance of the
Planck constant in physics would then have to be explained.  Its
origin should be in a scale, such as perhaps the original scale of a
universe, hidden in   gravity theory, or in some more fundamental
framework at small distance scale\footnote{One shoud keep  in
mind that these conclusions rest on the assumption (see
footnote~2) that classical general relativity remains valid up to
the Planck scale. The existence of some intermediate fundamental
scale could open  different perspectives.}.

\end{document}